\documentclass[aps,twocolumn,showpacs,preprintnumbers,floats]{revtex4}\def\@cite#1#2{\textsuperscript{[{#1\if@tempswa , #2\fi}]}}

\usepackage{graphicx}
\usepackage{amsmath}
\usepackage{amsfonts}
\usepackage{amssymb}
\usepackage{color}
\usepackage{subfigure}
\usepackage{epsfig}
\usepackage{morefloats}
\usepackage{multirow}
\usepackage{graphicx,booktabs}
\usepackage{mathrsfs}
\usepackage{txfonts}
\usepackage{indentfirst}
\usepackage{graphicx,booktabs}

\usepackage{longtable,lscape}

\newcommand{\be}{\begin{equation}}
\newcommand{\ee}{\end{equation}}
\newcommand{\bea}{\begin{eqnarray}}
\newcommand{\eea}{\end{eqnarray}}
\newcommand{\bean}{\begin{eqnarray*}}
\newcommand{\eean}{\end{eqnarray*}}

\newcommand{\gapproxeq}{\lower
.7ex\hbox{$\;\stackrel{\textstyle >}{\sim}\;$}}
\newcommand{\lapproxeq}{\lower
.7ex\hbox{$\;\stackrel{\textstyle <}{\sim}\;$}}

\begin{document}

\title{\textbf{ Canonical interpretation of $Y(10750)$  and $\Upsilon(10860)$ in the $\Upsilon$ family }}
\author{
 Qi Li, Ming-Sheng Liu, Qi-Fang L\"{u} \footnote {E-mail: lvqifang@hunnu.edu.cn}, Long-Cheng Gui \footnote{E-mail: guilongcheng@hunnu.edu.cn}, and Xian-Hui Zhong
\footnote {E-mail: zhongxh@hunnu.edu.cn} }  \affiliation{ 1) Department
of Physics, Hunan Normal University,  Changsha 410081, China }

\affiliation{ 2) Synergetic Innovation
Center for Quantum Effects and Applications (SICQEA), Changsha 410081,China}

\affiliation{ 3) Key Laboratory of
Low-Dimensional Quantum Structures and Quantum Control of Ministry
of Education, Changsha 410081, China}


\begin{abstract}

Inspired by the new resonance $Y(10750)$, we calculate the masses and two-body OZI-allowed strong decays of the higher vector bottomonium sates within both screened and linear potential models. We discuss the possibilities of $\Upsilon(10860)$ and $Y(10750)$
as mixed states via the $S-D$ mixing. Our results suggest that $Y(10750)$
and $\Upsilon(10860)$ might be explained as mixed states between $5S$- and $4D$-wave vector $b\bar{b}$
states. The $Y(10750)$ and $\Upsilon(10860)$ resonances may correspond to the mixed states dominated by the $4D$- and $5S$-wave components, respectively. The mass and the strong decay behaviors of the $\Upsilon(11020)$ resonance are consistent with the assignment of the $\Upsilon(6S)$ state in the potential models.
\end{abstract}

\maketitle

\section{Introduction}

Very recently, the Belle Collaboration reported a new measurement of the $e^+e^-\to \Upsilon(nS)\pi^+\pi^-\ (n=1,2,3)$
cross sections at energies from 10.52 to 11.02 GeV using data collected with the Belle detector
at the KEKB asymmetric-energy $e^+e^-$ collider~\cite{Abdesselam:2019gth}. Besides two old vector states $\Upsilon(10860)$ and $\Upsilon(11020)$, a new resonance near 10.75 GeV, i.e. $Y(10750)$ as named in Ref.~\cite{Wang:2019veq}, was obviously found in the cross sections.
The Breit-Wigner mass and width of this new structure are found to be $M=(10752.7\pm 5.9^{+0.7}_{-1.1})$ MeV
and $\Gamma=(35.5^{+17.6}_{-11.3}$$^{+3.9}_{-3.3})$ MeV. The production processes indicate that the spin-parity numbers of these three states appearing in the cross sections should be $J^{PC}=1^{--}$. It is a great challenge for our understanding these states with the conventional
$S$-, and $D$-wave bottomonium ($b\bar{b}$) states in the potential models.

The general consensus is that $\Upsilon(10860)$ and $\Upsilon(11020)$
correspond to the $S$-wave vector $b\bar{b}$ states $\Upsilon(5S)$ and $\Upsilon(6S)$, respectively~\cite{Chao:2009,Eichten:2007qx,Ebert:2011jc,Meng:2007tk,Godfrey:2015dia,Segovia:2016xqb,Deng:2016ktl,Wang:2018rjg,Ferretti:2013vua,Wei-Zhao:2013sta}.
However, if one assigns $\Upsilon(10860)$ to $\Upsilon(5S)$, we should face several problems, such as
(i) the mass of $\Upsilon(5S)$ from the recent potential model
calculations is about $70-90$ MeV lower than the observed value of $\Upsilon(10860)$~\cite{Deng:2016ktl,Wang:2018rjg,Wei-Zhao:2013sta,Godfrey:2015dia,Segovia:2016xqb}; (ii)
the mass splittings $m[\Upsilon(5S)-\Upsilon(4S)]_{\mathrm{th}}\simeq 210$ MeV and $m[\Upsilon(6S)-\Upsilon(5S)]_{\mathrm{th}}\simeq 180$ MeV
predicted within various potential models~\cite{Deng:2016ktl,Wang:2018rjg,Wei-Zhao:2013sta,Godfrey:2015dia,Segovia:2016xqb} are inconsistent with the observations $m[\Upsilon(5S)-\Upsilon(4S)]_{\mathrm{exp}}\simeq 306$ MeV and $m[\Upsilon(6S)-\Upsilon(5S)]_{\mathrm{exp}}\simeq 115$ MeV.

For the newly observed $Y(10750)$, we also meet several problems if we explain it with a pure
$S$-, or $D$-wave vector bottomonium state. According to the predictions in potential models, the $Y(10750)$ resonance lies between the vector states $\Upsilon(5S)$ and $\Upsilon_{1}(3D)$~\cite{Wang:2018rjg,Ferretti:2013vua,Godfrey:2015dia,Segovia:2016xqb,Deng:2016ktl}.
 Thus, if the $Y(10750)$ resonance correspond to a pure vector $b\bar{b}$ state, it should be assigned to either $\Upsilon(5S)$ or $\Upsilon_{1}(3D)$. However, if one assigns the $Y(10750)$ resonance to $\Upsilon(5S)$, we should meet a problem at once: how do we assign the $\Upsilon(10860)$ in the bottomonium family? On the other hand, if one assigns the $Y(10750)$ resonance to the $\Upsilon_{1}(3D)$ state, we cannot understand the productions of $Y(10750)$ in the $e^+e^-\to \Upsilon(nS)\pi^+\pi^-\ (n=1,2,3)$ processes, where the production cross sections of the $D$-wave states should be strongly suppressed for their very tiny dielectron widths predicted in theory~\cite{Wang:2018rjg,Godfrey:2015dia,Badalian:2008ik,Badalian:2009bu}.

The above analysis indicates that both $\Upsilon(10860)$ and $Y(10750)$ cannot be simply explained with a pure $S$- or $D$-wave $b\bar{b}$ state with $J^{PC}=1^{--}$. Thus, in the literature these resonances were suggested to be exotic states, such as, compact tetraquarks~\cite{Wang:2019veq,Ali:2009es}, or hadrobottomonium~\cite{Dubynskiy:2008mq}. It should be emphasized that although $\Upsilon(10860)$ and $Y(10750)$ are not good candidates of a pure $S$- or $D$-wave $b\bar{b}$ states, we cannot exclude them as mixed states between the $S$- and $D$-wave vector $b\bar{b}$ states. In Refs.~\cite{Badalian:2009bu,Badalian:2008ik}, Badalian \emph{et al.} studied
the dielectron widths of the vector bottomonium states, their results indicate that there might be sizeable $S-D$
mixing between the $nS$- and $(n-1)D$-wave ($n\geq 4$) vector states. If there is $S-D$ mixing indeed,
the masses of the pure $S$- and $D$-wave states should be shifted to the physical
states by some interactions. Then we may overcome the mass puzzles
of the $\Upsilon(10860)$ as a pure $\Upsilon(5S)$ state. On the other hand, if there is $S-D$ mixing indeed,
the physical states might have sizeable components of both $S$- and $D$-wave states. Considering the $Y(10750)$
as a mixed state dominated by the $D$-wave component, we may explain the large production cross sections in the $e^+e^-\to \Upsilon(nS)\pi^+\pi^-\ (n=1,2,3)$ processes for its sizeable $S$-wave component. In fact, the $S-D$ mixing might also exist in the other meson spectra, such as the $\psi(3770)$ is suggested to be a $1^3D_1$ state with a small admixture
of $2^3S_1$ state in the $c\bar{c}$ family~\cite{Eichten:2007qx,Eichten:2004uh,Eichten:2005ga,Rosner:2004wy}, while the $D^*_J(2600)$ and $D^*_{s1}(2700)$ might be mixed states via the $2^3S_1-1^3D_1$ mixing in the $D$ and $D_s$ meson families, respectively~\cite{Close:2006gr,Chen:2016spr,Chen:2011rr,Li:2010vx,Zhong:2010vq,Li:2009qu,Zhong:2009sk}.

In this work, we discuss the possibilities of $\Upsilon(10860)$ and $Y(10750)$
as mixed states via the $S-D$ mixing. By analyzing the mass spectrum of higher vector
bottomonium states above the $B\bar{B}$ threshold within both screened and linear
potential models, and calculating their strong decays with the $^3P_0$ model, we suggest that $Y(10750)$
and $\Upsilon(10860)$ might correspond to the two mixed states between $5S$- and $4D$-wave vector $b\bar{b}$
states with a sizeable mixing angle. The $Y(10750)$ and $\Upsilon(10860)$ resonances could be the $S-D$ mixed states dominated by the $4D$- and
$5S$-wave components, respectively.

This paper is organized as follows. The mass spectrum of higher vector bottomonium is calculated in Sec.~\ref{spectrum}. The $^3P_0$ model is briefly introduced and strong decays the vector $b\bar{b}$ states are calculated in Sec.~\ref{Strongdecay}. Then, combining the mass and widths, we carry out a discussion about the properties of the $J^{PC}=1^{--}$ states $Y(10750)$, $\Upsilon(10860)$ and $\Upsilon(11020)$ in Sec.~\ref{DIS}.
Finally, a short summary is presented in Sec.~\ref{sum}.

\begin{figure*}[!htbp]
\begin{center}
\centering  \epsfxsize=16cm \epsfbox{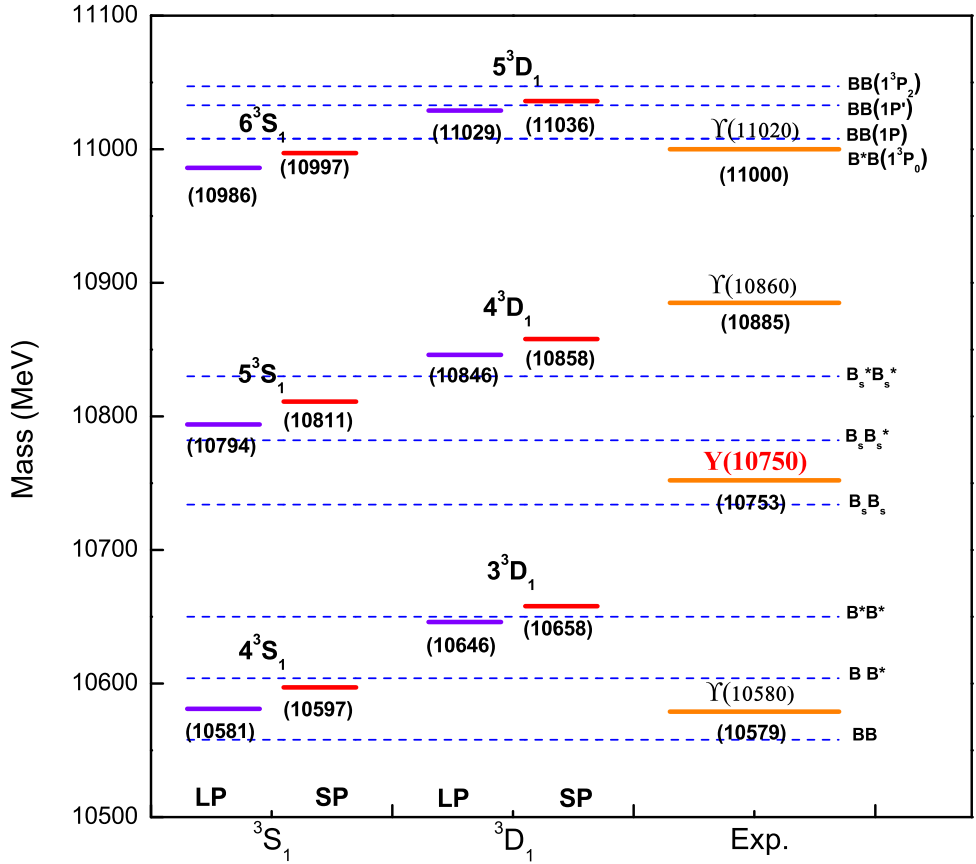}
\vspace{-1.5cm}\caption{The spectrum of the higher vector bottomonium above the $B\bar{B}$ threshold predicted within both screened potential (SP) and liner potential (LP) models. For a comparison, the experimental observations~\cite{Abdesselam:2019gth,Tanabashi:2018oca} are plotted in the figure as well.  } \label{mass}
\end{center}
\end{figure*}

\begin{table*}[htb]
\begin{center}
\caption{Predicted masses (MeV) of higher vector bottomonium states with both the linear potential (LP) and screened potential (SP) models. For a comparison, the results from recent works and experimental observations are also listed.}\label{tabs}
\begin{tabular}{ccccccccc}

\hline\hline
~~State       ~~&$n ^{2S+1}L_{J}$             ~~&LP    ~~&SP          ~~&Ref.~\cite{Wang:2018rjg}  ~~&Ref.~\cite{Godfrey:2015dia}    ~~&Ref.~\cite{Segovia:2016xqb} ~~&Exp. \\
\hline
$\Upsilon(4S)$    &$4 ^3S_{1}$       &10581   &10597         &10612                       &10635                            &10607    &$10579$ ~\cite{Tanabashi:2018oca}                    \\
$\Upsilon(5S)$    &$5 ^3S_{1}$        &10794   &10811         &10822                       &10878                            &10818     &$10885$~\cite{Abdesselam:2019gth}                  \\
$\Upsilon(6S)$    &$6 ^3S_{1}$        &10986   &10997         &11001                       &11102                            &10995     &$11000$~\cite{Abdesselam:2019gth}              \\
$\Upsilon_1(3D)$  &$3 ^3D_{1}$                 &10646   &10658         &10675                       &10698                            &10653     &$\cdots$                   \\
$\Upsilon_1(4D)$  &$4 ^3D_{1}$                  &10846   &10858         &10871                       &10928                            &10853     &$\cdots$                  \\
$\Upsilon_1(5D)$  &$5 ^3D_{1}$                  &11029   &11036         &11041                       &$\cdots$                         &11023   &$\cdots$                    \\
\hline\hline

\end{tabular}
\end{center}
\end{table*}

\section{mass spectrum}\label{spectrum}

The mass spectrum of bottomonium has been calculated in our previous works~\cite{Deng:2016ktl,Liu:2019zuc}
within the widely used linear potential model~\cite{Eichten:1974af,Eichten:1978tg,Godfrey:1985xj,Eichten:2007qx,Barnes:2005pb,Ferretti:2013vua}
and screened potential model~\cite{Li:2009zu,Chao:2009,ChaoKT93,Liu:2011yp,Lu:2016mbb}.
In these potential models, the effective potential of spin-independent term $V(r)$ is regarded as
the sum of Lorentz vector $V_V(r)$ and Lorentz scalar $V_S(r)$
contributions~\cite{Eichten:2007qx}, i.e.,
\begin{eqnarray}\label{vr}
V(r)=V_V(r)+V_S(r).
\end{eqnarray}
The Lorentz vector potential $V_V(r)$ can be written as the standard color
Coulomb form
\begin{eqnarray}\label{vv}
V_V(r)=-\frac{4}{3}\frac{\alpha_s}{r}.
\end{eqnarray}
The Lorentz scalar potential $V_S(r)$ might be taken as a simple form with a linear potential~\cite{Eichten:1974af,Eichten:1978tg,Eichten:2007qx,Barnes:2005pb,Ferretti:2013vua} or the screened potential as suggested in Refs.~\cite{Li:2009zu,Chao:2009,ChaoKT93,Liu:2011yp,Lu:2016mbb}, i.e.,
\begin{eqnarray}\label{vs}
V_S(r)=
\begin{cases}
br&\text{Linear potential}\\
\frac{b(1-e^{-\mu r})}{\mu}&\text{Screened potential}.
\end{cases}
\end{eqnarray}

Furthermore, we include three spin-dependent terms as follows. For the
spin-spin contact hyperfine potential, we take~\cite{Barnes:2005pb}
\begin{eqnarray}\label{ss}
H_{SS}= \frac{32\pi\alpha_s}{9m_b^2}\tilde{\delta}_\sigma(r)\mathbf{S}_b\cdot \mathbf{S}_{\bar{b}},
\end{eqnarray}
where $\mathbf{S}_b$ and $\mathbf{S}_{\bar{b}}$  are spin matrices acting on the spins
of the quark and antiquark. We take $\tilde{\delta}_\sigma(r)=(\sigma/\sqrt{\pi})^3
e^{-\sigma^2r^2}$ as in Ref.~\cite{Barnes:2005pb}.
For the spin-orbit and the tensor terms, we adopt~\cite{Eichten:2007qx}:
\begin{eqnarray}\label{sl}
H_{SL}= \frac{1}{2m_b^2r}\left(3\frac{dV_V}{dr}-\frac{dV_s}{dr}\right)\mathbf{L}\cdot \mathbf{S},
\end{eqnarray}
and
\begin{eqnarray}\label{t}
H_{T}= \frac{1}{12m_b^2}\left(\frac{1}{r}\frac{dV_V}{dr}-\frac{d^2V_V}{dr^2}\right)S_T,
\end{eqnarray}
where $\mathbf{L}$ is the relative orbital angular momentum of $b$ and $\bar{b}$ quarks,
$\mathbf{S}=\mathbf{S}_b+\mathbf{S}_{\bar{b}}$ is the total quark spin, and the
spin tensor $S_T$ is defined by~\cite{Eichten:2007qx}
\begin{eqnarray}\label{st}
S_T= 6\frac{\mathbf{S}\cdot \mathbf{r}\mathbf{S}\cdot \mathbf{r}}{r^2}-2\mathbf{S}^2.
\end{eqnarray}
If the linear potential is adopted, four parameters ($\alpha_s$, $b$, $m_b$, $\sigma$) in the above equations should be determined,
while if the screened potential is adopted, five parameters ($\alpha_s$, $b$, $m_b$, $\sigma$, $\mu$) should be
determined.

We solve the radial Schr\"{o}dinger equation by using the three-point difference central method~\cite{Haicai}.
With this method, one can reasonably include the corrections from these spin-dependent potentials to both the mass and wave
function of a meson state. The details of the numerical method can be found in our previous works~\cite{Deng:2016ktl,Deng:2016stx}.
With the same parameter sets determined in our previous works, we calculated the masses of the vector $b\bar{b}$ states, $nS$ ($n=4,5,6$), $nD$ ($n=3,4,5$), above the $B\bar{B}$ threshold within both the linear and screened potential models.

The calculated bottomonium masses are listed in Table~\ref{tabs} and also shown in Fig.~\ref{mass}. It is found that both the linear and screened potential models give a similar prediction of the masses. Considering the $\Upsilon(10580)$ and $\Upsilon(11020)$ resonances as the $\Upsilon(4S)$ and $\Upsilon(6S)$ states, respectively, their observed masses are in good agreement with the potential model predictions.
However, considering the $\Upsilon(10860)$ as the $\Upsilon(5S)$ state, the observed mass is obviously (about $70-90$ MeV) larger than the model predictions. Fig.~\ref{mass} shows that the newly observed state $Y(10750)$ lies about 100 MeV above $\Upsilon_{1}(3D)$, while about $40-50$ MeV below $\Upsilon(5S)$.


\section{strong decays}\label{Strongdecay}


In this section, we use the $^3P_0$ model~\cite{Micu:1968mk,LeYaouanc:1972vsx,LeYaouanc:1973ldf} to evaluate the Okubo-Zweig-Iizuka (OZI)  allowed two-body strong decays of the vector bottomonium. In this model, it assumes that the vacuum produces a light quark-antiquark pair with the quantum
number $0^{++}$ and the bottomonium decay takes place though the rearrangement of the four quarks.
The transition operator $\hat{T}$ can be written as
\begin{eqnarray}
    \hat{T} & = & -3 \gamma \sqrt{96 \pi} \sum_{m}^{} \langle 1 m 1 -m| 0 0 \rangle \int_{}^{} d\mathbf{p}_3 d\mathbf{p}_4 \delta^3 (\mathbf{p}_3 + \mathbf{p}_4) \nonumber\\
      & \times &  \mathcal{Y}_1^m \left(\frac{\mathbf{p}_3 - \mathbf{p}_4}{2}\right)  \chi_{1-m}^{34}  \phi_0^{34} \omega_0^{34} b_{3i}^\dagger (\mathbf{p}_3) d_{4j}^\dagger (\mathbf{p}_4) \ ,
\end{eqnarray}
where $\gamma$ is a dimensionless constant that denotes the strength of the quark-antiquark pair creation with
momentum $\mathbf{p}_3$ and $\mathbf{p}_4$ from vacuum; $b_{3i}^\dagger (\mathbf{p}_3)$ and $d_{4j}^\dagger(\mathbf{p}_4)$ are the creation operators for the quark and antiquark, respectively; the subscriptions, $i$ and $j$, are the SU(3)-color indices of the created quark and anti-quark;
$\phi_0^{34}=(u\bar u +d\bar d +s \bar s)/\sqrt 3$ and $\omega_{0}^{34}=\frac{1}{\sqrt{3}} \delta_{ij}$ correspond to flavor and
color singlets, respectively; $\chi_{{1,-m}}^{34}$ is a spin triplet
state; and $\mathcal{Y}_{\ell m}(\mathbf{k})\equiv
|\mathbf{k}|^{\ell}Y_{\ell m}(\theta_{\mathbf{k}},\phi_{\mathbf{k}})$ is the
$\ell$-th solid harmonic polynomial.

For an OZI allowed two-body strong decay process $A\to B+C$, the helicity amplitude
$\mathcal{M}^{M_{J_A}M_{J_B} M_{J_C}}(\mathbf{P})$ can be calculated as follow
\begin{eqnarray}
\langle BC|T| A\rangle=\delta(\mathbf{P}_A-\mathbf{P}_B-\mathbf{P}_C)\mathcal{M}^{M_{J_A}M_{J_B} M_{J_C}}(\mathbf{P}).
\end{eqnarray}
With the Jacob-Wick formula~\cite{Jacob:1959at},
the helicity amplitudes
$\mathcal{M}^{M_{J_A}M_{J_B} M_{J_C}}(\mathbf{P})$ can be converted to the partial wave amplitudes $\mathcal{M}^{JL}$ via
\begin{equation}\label{eq4}
\begin{aligned}
&
{\mathcal{M}}^{J L}(A\rightarrow BC) = \frac{\sqrt{4\pi (2 L+1)}}{2 J_A+1} \!\! \sum_{M_{J_B},M_{J_C}} \langle L 0 J M_{J_A}|J_A M_{J_A}\rangle \\
& \hspace{1cm}
\times  \langle J_B M_{J_B} J_C M_{J_C} | J M_{J_A} \rangle \mathcal{M}^{M_{J_A} M_{J_B} M_{J_C}}({\textbf{P}}).
\end{aligned}
\end{equation}
In the above equations, ($J_{A}$, $J_{B}$ and $J_{C}$), ($L_A$, $L_B$ and $L_C$) and ($S_A$, $S_B$ and $S_C$) are the quantum numbers of the total angular momenta, orbital angular momenta and total spin for hadrons $A,B,C$, respectively; $M_{J_A}=M_{J_B}+M_{J_C}$ ,\;$\mathbf{J}\equiv \mathbf{J}_B+\mathbf{J}_C$ and $\mathbf{J}_{A} \equiv \mathbf{J}_{B}+\mathbf{J}_C+\mathbf{L}$.
In the c.m. frame of hadron $A$, the momenta $\mathbf{P}_B$ and $\mathbf{P}_C$ of mesons $B$ and $C$ satisfy
$\mathbf{P}_B=-\mathbf{P}_C\equiv \mathbf{P}$.

To partly remedy the inadequacy of the nonrelativistic
wave function as the momentum $\boldsymbol{P}$ increases, a Lorentz boost factor $\gamma_f$ is introduced into the decay
amplitudes~\cite{Gui:2018rvv},
\begin{eqnarray}
{\mathcal{M}} \to \gamma_f {\mathcal{M}(\gamma_f \boldsymbol{P})}.
\end{eqnarray}
where $\gamma_f = M_B/E_B$. In the decays with small phase space, the three momenta $\boldsymbol{P}$
carried by the final state mesons and corrections from the Lorentz boost are relatively small, while the relativistic effects may be essential for the decay channels with larger phase space.

Finally, the partial width $A\to B+C$ can be given by
\begin{eqnarray}
\Gamma = 2\pi |\textbf{P}| \frac{E_BE_C}{M_A}\sum_{JL}\Big
|\mathcal{M}^{J L}\Big|^2,\label{de}
\end{eqnarray}
where $M_A$ is the mass of the initial hadron $A$, while $E_B$ and $E_C$ stand for the energies of
final hadrons $B$ and $C$, respectively. The details of the formula of the $^3P_0$ model can be found in Refs~\cite{Gui:2018rvv,Li:2019tbn}.

In the calculations, the wave functions of the initial vector bottomonium states are taken from our quark model predictions.
Furthermore, we need the wave functions of the final hadrons, i.e.,
the $B^{(*)}$ and $B_s^{(*)}$ mesons and some of their excitations,
which are adopted from the quark model predictions of Refs.~\cite{Lu:2016bbk,Li:2010vx}. The masses of the final hadron states in
the decay processes are adopted from the Particle Data Group~\cite{Tanabashi:2018oca} if there are data, while if there are no observations we adopt the predicted values from Refs.~\cite{Lu:2016bbk,Li:2010vx}. The quark pair creation strength is determined to be $\gamma = 0.232$ by reproducing the measured width $\Upsilon(10580) \to B \bar B = 20.5~\rm{MeV}$~\cite{Tanabashi:2018oca} with the wave function calculated from the screened potential model. The $\gamma$ determined in this work is also close to the values $0.217/0.234$ adopted in the study of the strong decays of excited charmonium states~\cite{Gui:2018rvv}.
The strong decay properties for the vector bottomonium are presented in Table~\ref{decay}. It is found that both the linear and screened potential models give similar predictions for the strong decay properties of the bottomonium states.

\begin{figure}[!htbp]
\begin{center}
\centering  \epsfxsize=9.2 cm \epsfbox{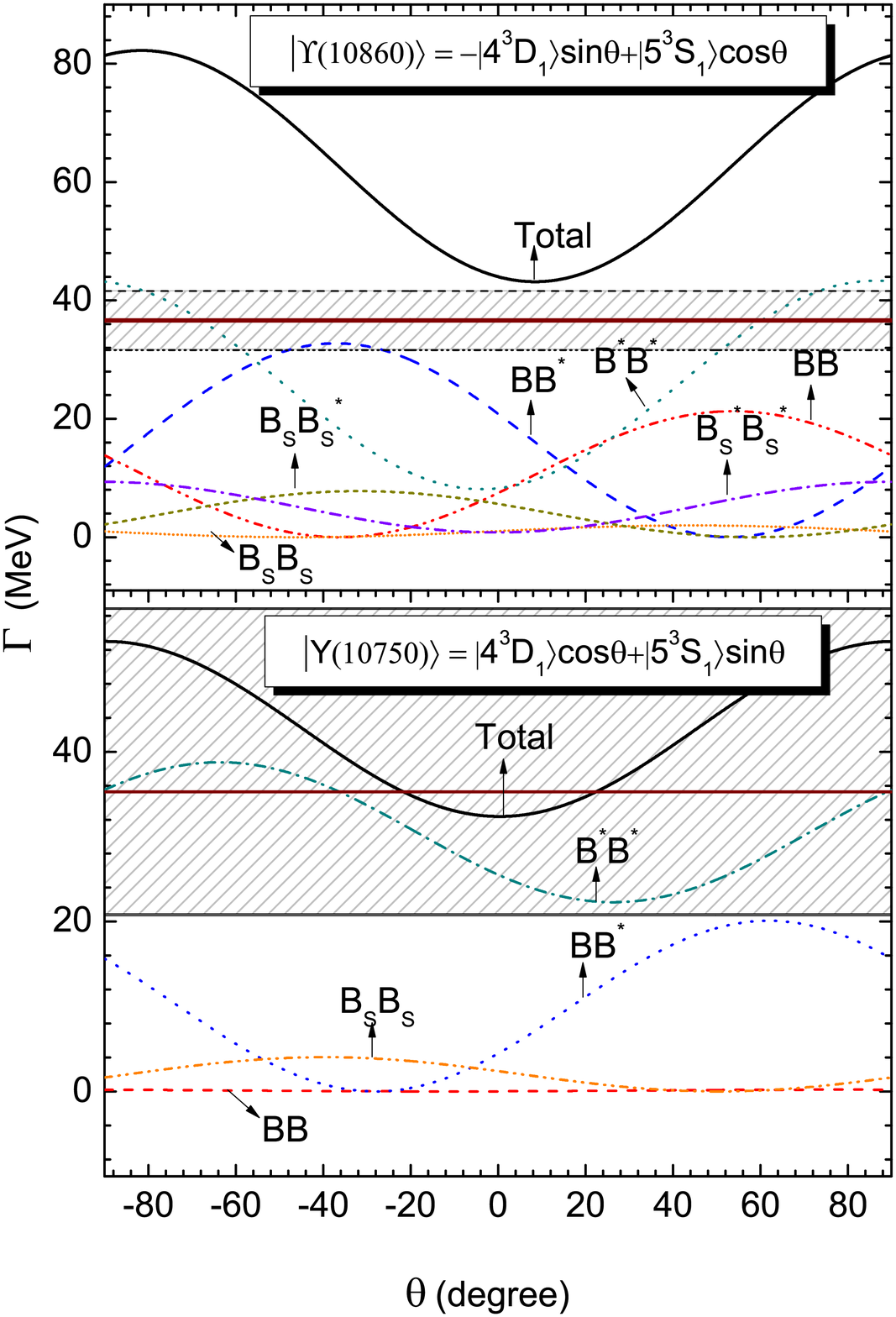}
\vspace{-1.2 cm}\caption{Strong decays of $Y(10750)$ and $\Upsilon(10860)$ versus the mixing angle $\theta$ within screened potential model.} \label{5s4d}
\end{center}
\end{figure}

\begin{figure}[htbp]
\begin{center}
\centering  \epsfxsize=9.2cm \epsfbox{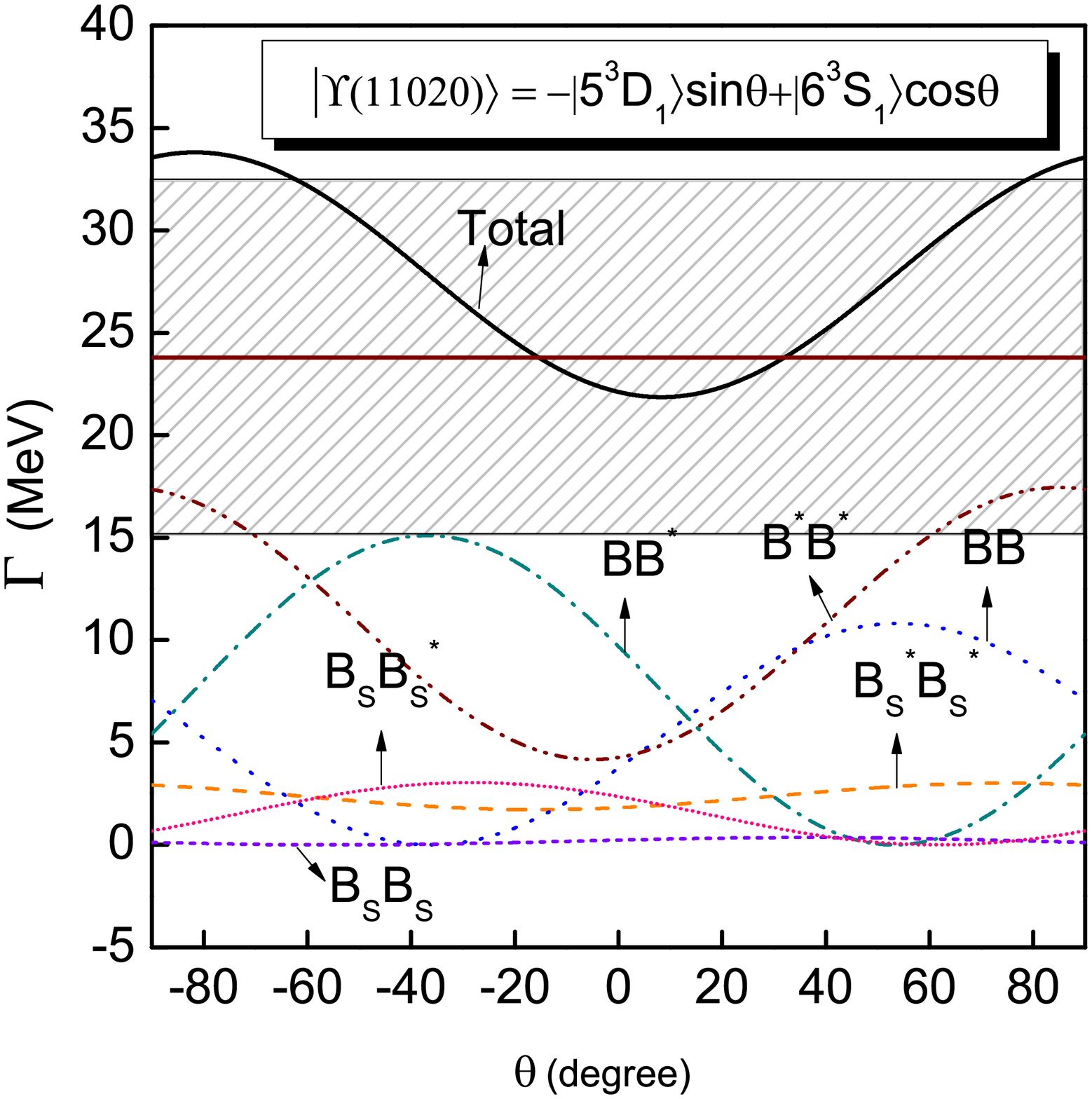}
\vspace{-0.8cm}\caption{Strong decay of $\Upsilon(11020)$ versus the mixing angle $\theta$ within screened potential model.} \label{6s5d}
\end{center}
\end{figure}

\begin{figure}[htbp]
\begin{center}
\centering  \epsfxsize=9.2cm \epsfbox{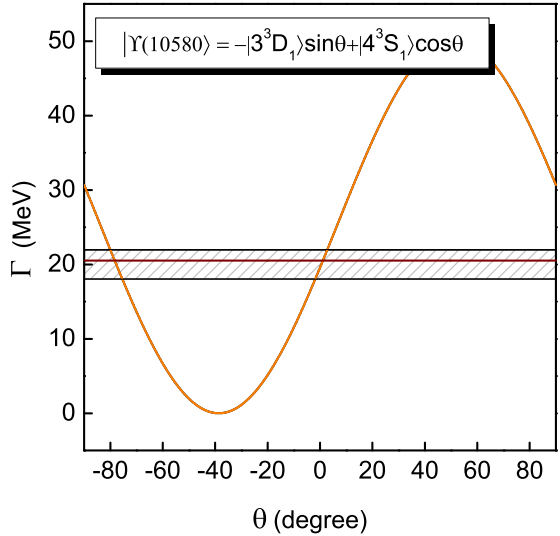}
\vspace{-0.8cm}\caption{Strong decay of $\Upsilon(10580)$ versus the mixing angle $\theta$ within screened potential model.} \label{4s3d}
\end{center}
\end{figure}

\section{discussions}\label{DIS}

\subsection{$Y(10750)$ and $\Upsilon(10860)$}

For the $\Upsilon(10860)$ resonance, the mass and spin-parity numbers indicate that it may be a candidate of the conventional $\Upsilon_1(4D)$.
However, with this assignment it is found that the total decay width, $\sim$81 MeV (see Table~\ref{decay4D}), is about a factor of 2 larger than the measured width $\sim 37~\rm{MeV}$~\cite{Abdesselam:2019gth}. Furthermore, assigning $\Upsilon(10860)$ as $\Upsilon_1(4D)$ we will meet a problem in the explanation of its productions in the $e^+e^-\to \Upsilon(nS)\pi^+\pi^-\ (n=1,2,3)$, because the production rates of $D$-wave states should be strongly suppressed for their tiny dielectron widths~\cite{Wang:2018rjg,Godfrey:2015dia,Segovia:2016xqb,Badalian:2008ik,Badalian:2009bu}.
The $\Upsilon(10860)$ resonance is often explained as the $\Upsilon(5S)$ state in the literature~\cite{Wang:2018rjg,Ferretti:2013vua,Godfrey:2015dia,Segovia:2016xqb}, with this assignment, the total width is predicted to be $\sim 44$ MeV (see Table~\ref{decay5S}), which is consistent with the data. However, if we assign $\Upsilon(10860)$ to $\Upsilon(5S)$ we should face some problems, for example,
(i) the predicted mass of $\Upsilon(5S)$ state is about 70 MeV lower than that of $\Upsilon(10860)$; (ii) moreover,
the mass splittings $m[\Upsilon(5S)-\Upsilon(4S)]_{\mathrm{th}}\simeq 210$ MeV and $m[\Upsilon(6S)-\Upsilon(5S)]_{\mathrm{th}}\simeq 180$ MeV
predicted within various potential models (see Table~\ref{tabs}) are inconsistent with the observations
$m[\Upsilon(5S)-\Upsilon(4S)]_{\mathrm{exp}}\simeq 306$ MeV and $m[\Upsilon(6S)-\Upsilon(5S)]_{\mathrm{exp}}\simeq 115$ MeV~\cite{Tanabashi:2018oca}.

For the new structure $Y(10750)$ observed at Belle, from Figure~\ref{mass} one finds that it lies between the vector states $\Upsilon(5S)$ and $\Upsilon_{1}(3D)$. The predicted mass of $\Upsilon_{1}(3D)$ state is about 100 MeV lower than that of $Y(10750)$. If one assigns the $Y(10750)$ resonance to the $\Upsilon_{1}(3D)$ state, we cannot understand its production rates in the $e^+e^-\to \Upsilon(nS)\pi^+\pi^-\ (n=1,2,3)$ processes. The production cross sections of the $D$-wave states should be strongly suppressed for their very tiny dielectron widths predicted in theory~\cite{Wang:2018rjg,Godfrey:2015dia,Segovia:2016xqb,Badalian:2008ik,Badalian:2009bu}.
Thus, the explanation of the $Y(10750)$ resonance as the $\Upsilon_{1}(3D)$ state should be excluded.
On the other hand, if one assigns the $Y(10750)$ resonance to $\Upsilon(5S)$, it is found that the decays of $Y(10750)$ are dominated by the $B^*B^*$ channel, and the decay width is predicted to be $\Gamma \sim 53$ MeV, which is close to the measured value $35.5^{+17.6+3.9}_{-11.3-3.3}~\rm{MeV}$ at Belle~\cite{Abdesselam:2019gth}. However, we will meet a problem that there are no $S$-wave vector states to be assigned to $\Upsilon(10860)$, then we cannot understand largest production rates of $\Upsilon(10860)$ in the $e^+e^-\to \Upsilon(nS)\pi^+\pi^-\ (n=1,2,3)$ processes.

Since it is difficult to assign the $Y(10750)$ and $\Upsilon(10860)$  as the pure $S$- and $D$-wave $b\bar{b}$ states simultaneously, we consider the possibilities of $Y(10750)$ and $\Upsilon(10860)$ as the $\Upsilon(5S)$-$\Upsilon_1(4D)$ mixed states with the following mixing scheme
\begin{equation}\label{eqmx}
\left(
  \begin{array}{c}
   Y(10750)\\
   \Upsilon(10860)\\
  \end{array}\right)=
  \left(
  \begin{array}{cc}
   \cos\theta  &\sin\theta \\
  -\sin\theta  &\cos\theta \\
  \end{array}
\right)
\left(
  \begin{array}{c}
  \Upsilon_1(4D)\\
  \Upsilon(5S)\\
  \end{array}\right).
\end{equation}
The decay widths of $Y(10750)$ and $\Upsilon(10860)$ versus the mixing angle $\theta$ are presented in Fig.~\ref{5s4d}.
In Refs.~\cite{Badalian:2008ik,Badalian:2009bu}, Badalian \emph{et al.} studied
the dielectron widths of the vector bottomonium states, their results indicate that there might be sizeable $S-D$
mixing between the $nS$- and $(n-1)D$-wave ($n\geq 4$) vector states with a mixing angle $\sim 27^\circ$.
With this mixing angle, the decay widths of $Y(10750)$ and $\Upsilon(10860)$ can be reasonably understood (see Fig.~\ref{5s4d}).

The $S-D$ mixing mechanics can shift the masses of the pure states $\Upsilon(5S)$ and $\Upsilon_1(4D)$ to the physical states $Y(10750)$ and $\Upsilon(10860)$.
The Hamiltonian of the physical states is assumed to be
\begin{equation}
H=H_0+H_I,
\end{equation}
where $H$ contributes diagonal elements to the mass matrix, while $H_I$  could contribute non-diagonal elements to the mass matrix causing the $S-D$ mixing.
Then, the masses of $Y(10750)$ and $\Upsilon(10860)$ can be determined by
\begin{eqnarray}
M[\Upsilon(10860)]&=& \langle \Upsilon(10860) |H_0+H_I|\Upsilon(10860)\rangle,\\
M[Y(10750)]&=& \langle Y(10750) |H_0+H_I|Y(10750)\rangle.
\end{eqnarray}
Combining the mixing scheme defined in Eq.(\ref{eqmx}), one finds that
\begin{eqnarray}
M[\Upsilon(10860)]&=& M[\Upsilon_1(4D)]\sin^2\theta
      +M[\Upsilon(5S)]\cos^2\theta\nonumber\label{mix1}\\
      && -\Delta M_{SD} \sin (2\theta),\\
M[Y(10750)]&=&  M[\Upsilon_1(4D)]\cos^2\theta
      + M[\Upsilon(5S)]\sin^2\theta\nonumber\\
      &&+ \Delta M_{SD}\sin (2\theta),
\end{eqnarray}
where $M[\Upsilon(5S)]=\langle \Upsilon(5S)|H|\Upsilon(5S)\rangle$ and
$M[\Upsilon_1(4D)]=\langle \Upsilon_1(4D)|H|\Upsilon_1(4D)\rangle$ correspond to the masses
of the pure states $\Upsilon(5S)$ and $\Upsilon_1(4D)$, respectively, while
$\Delta M_{SD}=\langle \Upsilon_1(4D)|H_I|\Upsilon(5S)\rangle$ corresponds to
the non-diagonal element. Taking a sizeable mixing angle $\theta\sim 20-30^\circ$ and a negative value $\Delta M_{SD}\sim -100$ MeV in Eq.(\ref{mix1}), one finds that the physical masses of both $Y(10750)$ and $\Upsilon(10860)$ can be consistent with the experimental observations. Thus, with the $S-D$ mixing one may overcome the puzzle that the observed mass of $\Upsilon(10860)$ is obviously higher than the predicted masses of pure $\Upsilon(5S)$ in the potential models.

As a pure $\Upsilon(5S)$ state, the dielectron width of $\Upsilon(10860)$ is predicted to be $\Gamma_{ee}=0.348$ keV in a recent work~\cite{Wang:2018rjg}.
Combing it with the mixing angle $\theta \simeq 27^\circ$ suggested in Ref.~\cite{Badalian:2008ik,Badalian:2009bu} we obtain dielectron width $\Gamma_{ee}=0.28$ keV
for $\Upsilon(10860)$, which is consistent with the measured value $0.31\pm 0.07$ keV~\cite{Tanabashi:2018oca}.
As a mixed state containing sizeable $S$-wave component the dielectron width of $Y(10750)$ is estimated to
be $\sim 0.07$ keV. Neglecting the effect of phase space, one may predict the ratio between the production rates of $Y(10750)$ and $\Upsilon(10860)$ in the $e^+e^-\to \Upsilon(nS)\pi^+\pi^-\ (n=1,2,3)$ processes, i.e.,
\begin{eqnarray}
R\sim \frac{\Gamma_{ee}[Y(10750)]}{\Gamma_{ee}[\Upsilon(10860)]}\simeq \frac{1}{4},
\end{eqnarray}
which can explain the observations that production cross sections of $Y(10750)$ are comparable with
those of $\Upsilon(10860)$ in the $e^+e^-\to \Upsilon(nS)\pi^+\pi^-\ (n=1,2,3)$ processes.

From Fig.~\ref{5s4d}, one can see that the partial widths of the strong decay modes and the ratios between them for the $Y(10750)$ and $\Upsilon(10860)$ resonances are sensitive to the mixing angle. If taking the mixing angle as $\theta\sim 30^\circ$, the decays of $\Upsilon(10860)$ are dominated by the $B^*B^*$, $BB$ and $B_s^*B_s^*$ channels, while the decays of $Y(10750)$ might be governed by both the $B^*B^*$ and $BB^*$ channels. The large decay rates of $\Upsilon(10860)$ into the $B_s^*B_s^*$ channel can explain the observations at Belle~\cite{Abdesselam:2016tbc} that the $B_s^*B_s^*$ cross section shows a prominent $\Upsilon(10860)$ signal, while the $B_s^*B_s$ and $B_sB_s$ cross sections are relatively small and do not show any significant structures. From the Review of Particle Physics~\cite{Tanabashi:2018oca}, the branching ratios of $BB$, $BB^*$, $B^*B^*$, $B_sB_s$, $B_sB_s^*$, and $B_s^*B_s^*$ decay modes are $5.5\pm1.0\%$, $13.7\pm1.6\%$, $38.1\pm3.4\%$, $0.5\pm0.5\%$, $1.35\pm0.32\%$, and $17.6\pm2.7\%$, respectively. From Table~\ref{decay5S}, it can be seen that these branching ratios of $\Upsilon(10860)$ can be hardly described in the pure $\Upsilon(5S)$ interpretation, which is consistent with the analysis in Refs.~\cite{Godfrey:2015dia,Wang:2018rjg,Segovia:2016xqb}. With the mixing scheme, this problem can be partially overcame. Our results show that the $B^*B^*$ dominates in the non-strange final channels and the $B_s^*B_s^*$ is prominent in the strange decay modes, which is consistent with the experimental data. However, the predicted large $BB$ partial decay width is still in conflicting with the observations. More theoretical and experimental investigations are needed to clarify this puzzle.

The intermediate $B^*B^*$ meson loop may play an important role in the $S-D$ mixing between $\Upsilon(5S)$ and $\Upsilon_1(4D)$.
From Table~\ref{decay}, it is seen that both $\Upsilon(5S)$ and $\Upsilon_1(4D)$ states strongly couple to the $B^*B^*$ channel, thus, intermediate $B^*B^*$ meson loop may contribute a sizeable non-diagonal element to the mass matrix, which leads to the $S-D$ mixing.
It is interesting to find that the mixing mechanism of axial-vectors $D_{s1}(2460)$ and $D_{s1}(2536)$ has been studied via intermediate
hadron loops, e.g. $DK$, to which both states have strong couplings in Ref.~\cite{Wu:2011yb}. Also, the $S-D$ mixing scheme of $\psi(4S)$ and $\psi_1(2D)$ states via the meson loops are investigated in Ref.~\cite{Cao:2017lui}. Their results indicate that the intermediate
hadron loops as the mixing mechanism can lead to strong configuration mixing effects and obvious mass shifts for the physical states.

\begin{table*}[!htbp]
\caption{ Strong decay properties for the higher vector bottomonium states predicted within both LP and SP models. For a comparison, other results predicted in the recent works~\cite{Godfrey:2015dia,Wang:2018rjg,Segovia:2016xqb} are also listed in the same table.}\label{decay}
\begin{tabular}{cccccccccccccccccccccccccc}  \midrule[1.0pt]\midrule[1.0pt]
~~&\multirow{1}{*}{State}
~~&\multirow{2}{*}{Decay mode}
~~&\multicolumn{2}{c} {\underline{~~~~~~~~~~LP~~~~~~~~~~}}
~~&\multicolumn{2}{c} {\underline{~~~~~~~~~~SP~~~~~~~~~~}}
~~&\multicolumn{2}{c} {\underline{~~~~~~~~~~Ref.~\cite{Wang:2018rjg}~~~~~~~~~~}}
~~&\multicolumn{2}{c} {\underline{~~~~~~~~~~Ref.~\cite{Godfrey:2015dia}~~~~~~~~~~}}
~~&\multicolumn{2}{c} {\underline{~~~~~~~~~~Ref.~\cite{Segovia:2016xqb}~~~~~~~~~~}}
\\
~&(LP/SP)
~&
~&$\Gamma_{th}$(MeV) ~&$B_{r}(\%)$
~&$\Gamma_{th}$(MeV) ~&$B_{r}(\%)$
~&$\Gamma_{th}$(MeV) ~&$B_{r}(\%)$
~&$\Gamma_{th}$(MeV) ~&$B_{r}(\%)$
~&$\Gamma_{th}$(MeV) ~&$B_{r}(\%)$\\
\midrule[1.0pt]
~&$4 ^3S_{1}$	~&	$BB$	        ~&19.58       ~&100              ~&20.25	        ~&100		      ~&24.7    ~&100          ~&22.0       ~&100		 ~&20.59    ~&100	    \\
~&(10581/10597) ~&  Total	        ~&19.58       ~&100              ~&20.25		    ~&100	          ~&24.7    ~&100	       ~&22.0       ~&100	    ~&20.59    ~&100 	    \\
\\
~&$5 ^3S_{1}$	~&	$BB$	        ~&0.96        ~&2.23             ~&2.25	            ~&7.26	          ~&13.7    ~&30.0         ~&5.35       ~&19.5		 ~&6.22     ~&22.29 	    \\
~&(10794/10811)	~&	$BB^*$	        ~&1.76        ~&4.09             ~&$\approx0$       ~&$\approx0$	  ~&26.5    ~&58.1	       ~&16.6       ~&60.6		 ~&11.83    ~&42.41	    \\
~&		        ~&	$B^*B^*$	    ~&33.73       ~&78.37            ~&21.77		    ~&70.23	          ~&2.58    ~&5.66	       ~&2.42       ~&8.83		 ~&0.09     ~&0.32	    \\
~&		        ~&	$B_sB_s$	    ~&$\approx0$  ~&$\approx0$       ~&0.30		        ~&0.97	          ~&0.484	~&1.06	       ~&0.157      ~&0.573		~&0.96     ~&3.45	    \\
~&		        ~&	$B_s^*B_s$	    ~&6.59        ~&15.31            ~&6.68		        ~&21.55	          ~&1.49    ~&3.28	       ~&0.833      ~&3.04		 ~&1.15     ~&4.11	    \\
~&		        ~&	$B_s^*B_s^*$	~&            ~&                 ~&		            ~&	              ~&0.872   ~&1.91	       ~&2.00       ~&7.30		 ~&7.65     ~&27.42	    \\
~&	          	~&	Total           ~&43.04       ~&100              ~&31.00		    ~&100	          ~&45.6	~&100	       ~&27.4       ~&100		 ~&27.89    ~&100	    \\
\\
~&$6 ^3S_{1}$	~&	$BB$	        ~&3.22        ~&26.86            ~&3.80	            ~&18.90	          ~&7.81    ~&20.4         ~&1.32       ~&3.89		 ~&4.18      ~&5.28	    \\
~&(10986/10997)	~&	$BB^*$	        ~&5.69        ~&47.46            ~&9.02		        ~&44.85	          ~&16.5    ~&43.0	       ~&7.59       ~&22.4		 ~&15.49     ~&19.57	    \\
~&		        ~&	$B^*B^*$	    ~&0.44        ~&3.67             ~&3.13		        ~&15.56	          ~&4.43    ~&11.5	       ~&5.89       ~&17.4		 ~&11.87     ~&14.99	    \\
~&		        ~&	$B_sB_s$	    ~&0.38        ~&3.17             ~&0.28		        ~&1.39	          ~&0.101   ~&0.263	       ~&1.31       ~&0.00386   ~&0.07      ~&0.09	    \\
~&		        ~&	$B_s^*B_s$	    ~&1.94        ~&16.18            ~&2.38		        ~&11.83	          ~&0.780   ~&2.04	       ~&0.136      ~&0.401		~&1.50      ~&1.89	    \\
~&		        ~&	$B_s^*B_s^*$	~&0.32        ~&2.67             ~&1.50		        ~&7.46	          ~&0.448	~&1.17	       ~&0.310      ~&0.914		~&2.02      ~&2.56	    \\
~&		        ~&	$BB(1P)$	    ~&            ~&                 ~&		            ~&	              ~&8.27    ~&21.6	       ~&7.81       ~&23.0		 ~&40.08     ~&50.64	    \\
~&		        ~&	$BB(1P')$	    ~&            ~&                 ~&		            ~&	              ~&        ~&	           ~&10.8       ~&31.8		 ~&3.95      ~&4.98	    \\
~&	          	~&	Total           ~&12.4        ~&100              ~&20.11		    ~&100	          ~&38.3	~&100	       ~&33.9       ~&100		 ~&79.16     ~&100	    \\
\\
~&$3 ^3D_{1}$	~&	$BB$	        ~&$\approx0$  ~&$\approx0$       ~&0.95	            ~&3.41		      ~&5.47    ~&10.1         ~&23.8       ~&23.0		 ~&$\cdots$  ~&$\cdots$	\\
~&(10646/10658) ~&	$BB^*$	        ~&26.41       ~&100              ~&18.76		    ~&67.39	          ~&15.2    ~&28.1	       ~&0.245      ~&0.236		~&$\cdots$  ~&$\cdots$	\\
~&		        ~&	$B^*B^*$	    ~&            ~&                 ~&	8.13	       ~&29.20	              ~&33.4	~&61.8	       ~&79.5       ~&76.7		 ~&$\cdots$  ~&$\cdots$	\\
~&	          	~&	Total           ~&26.41       ~&100              ~&27.84		    ~&100	          ~&54.1	~&100	       ~&103.6      ~&100		 ~&$\cdots$  ~&$\cdots$	\\
\\
~&$4^3D_{1}$	~&	$BB$	        ~&11.82       ~&30.17            ~&12.84	       ~&23.87	          ~&27.4    ~&31.4         ~&3.85       ~&5.36		 ~&$\cdots$  ~&$\cdots$	\\
~&(10846/10858) ~&	$BB^*$	        ~&5.55        ~&14.17            ~&7.45		       ~&13.85	          ~&15.1	~&17.3	       ~&14.0       ~&19.5		 ~&$\cdots$  ~&$\cdots$	\\
~&		        ~&	$B^*B^*$	    ~&18.25       ~&46.58            ~&27.28		   ~&50.71	          ~&42.1	~&48.3	       ~&50.6       ~&70.5		 ~&$\cdots$  ~&$\cdots$	\\
~&		        ~&	$B_sB_s$	    ~&1.67        ~&4.3              ~&1.86		       ~&3.48	          ~&0.560	~&0.642	       ~&0.101      ~&0.141		~&$\cdots$  ~&$\cdots$	\\
~&		        ~&	$B_s^*B_s$	    ~&0.05        ~&0.13             ~&0.62		       ~&1.15	          ~&0.360	~&0.412	       ~&0.332      ~&0.462		~&$\cdots$  ~&$\cdots$	\\
~&		        ~&	$B_s^*B_s^*$	~&1.84        ~&4.70             ~&3.75		       ~&6.97	          ~&1.66	~&1.91	       ~&2.94       ~&4.09		 ~&$\cdots$  ~&$\cdots$	\\
~&	          	~&	Total           ~&39.18       ~&100              ~&53.8		       ~&100	          ~&87.2	~&100	       ~&71.8       ~&100		 ~&$\cdots$  ~&$\cdots$	\\
\\
~&$5^3D_{1}$	~&	$BB$	         ~&8.28       ~&14.27            ~&8.13	           ~&10.39	          ~&20.0    ~&16.4         ~&$\cdots$	 ~&$\cdots$  ~&$\cdots$  ~&$\cdots$	\\
~&(11029/11036) ~&	$BB^*$	         ~&7.89       ~&13.60            ~&8.49		       ~&10.85	          ~&19.3    ~&15.8	       ~&$\cdots$	 ~&$\cdots$	~&$\cdots$  ~&$\cdots$	\\
~&		        ~&	$B^*B^*$	     ~&24.08      ~&41.51            ~&27.41		   ~&35.02	          ~&47.1    ~&38.7	       ~&$\cdots$	 ~&$\cdots$	~&$\cdots$  ~&$\cdots$	\\
~&		        ~&	$B_sB_s$	     ~&$\approx0$ ~&$\approx0$       ~&0.02		       ~&0.03	          ~&0.0235	~&0.0193	   ~&$\cdots$   ~&$\cdots$	~&$\cdots$  ~&$\cdots$	\\
~&		        ~&	$B_s^*B_s$	     ~&0.44       ~&0.76             ~&0.36		       ~&0.46	          ~&0.103	~&0.0844	   ~&$\cdots$   ~&$\cdots$	~&$\cdots$  ~&$\cdots$	\\
~&		        ~&	$B_s^*B_s^*$	 ~&2.91       ~&5.02             ~&2.60		       ~&3.32	          ~&0.798   ~&0.656	       ~&$\cdots$   ~&$\cdots$	~&$\cdots$  ~&$\cdots$	\\
~&		        ~&	$B^*B(1^3P_0)$	 ~&7.92       ~&13.65            ~&10.08		   ~&12.88	          ~&3.02	~&2.48	       ~&$\cdots$	 ~&$\cdots$  ~&$\cdots$  ~&$\cdots$	\\
~&		        ~&	$BB(1P)$	     ~&6.49       ~&11.19            ~&13.56	       ~&17.33	          ~&4.08	~&3.35	       ~&$\cdots$	 ~&$\cdots$	~&$\cdots$  ~&$\cdots$	\\
~&		        ~&	$BB(1P')$	     ~&           ~&                 ~&7.61		       ~&9.72	          ~&18.1    ~&14.8	       ~&$\cdots$	 ~&$\cdots$	~&$\cdots$  ~&$\cdots$	\\
~&		        ~&	$BB(1^3P_2)$	 ~&           ~&                 ~&		           ~&	              ~&9.23	~&7.59	       ~&$\cdots$	 ~&$\cdots$	~&$\cdots$  ~&$\cdots$	\\
~&	          	~&	Total            ~&58.01      ~&100              ~&78.26		   ~&100	          ~&121.7	~&100	       ~&$\cdots$	 ~&$\cdots$	~&$\cdots$  ~&$\cdots$	\\
\midrule[1.0pt]\midrule[1.0pt]
\end{tabular}
\end{table*}

\begin{table}[!htbp]
\caption{Strong decays for the $\Upsilon_{1}(4D)$ which is assigned to be $Y(10750)$ or $\Upsilon(10860)$ within the screened potential model. }   \label{decay4D}
\begin{tabular}{cccccccccccccccccccccccccc}  \midrule[1.0pt]\midrule[1.0pt]
\multirow{2}{*}{State}
~~&\multirow{2}{*}{Mode}
~~&\multicolumn{2}{c} {\underline{~~~~~~~~~~$Y(10750)$~~~~~~~~~~}}
~~&\multicolumn{2}{c} {\underline{~~~~~~$\Upsilon(10860)$~~~~}}
\\

~&
~&$\Gamma_{th}$(MeV) ~&$B_{r}(\%)$
~&$\Gamma_{th}$(MeV) ~&$B_{r}(\%)$
\\
\midrule[1.0pt]
$4 ^3D_{1}$	~&	$BB$	        ~&$\approx0$           ~&$\approx0$                ~&13.82	        ~&16.98         \\
           	~&	$BB^*$	        ~&4.46       ~&13.77            ~&11.92            ~&14.65     	 \\
		        ~&	$B^*B^*$	    ~&25.53       ~&78.84            ~&43.17		    ~&53.04	         \\
		        ~&	$B_sB_s$	    ~&2.39        ~&7.38            ~&0.97		        ~&1.19           \\
		        ~&	$B_s^*B_s$	    ~&            ~&                 ~&2.16		        ~&2.65	         \\
		        ~&	$B_s^*B_s^*$	~&            ~&                 ~&9.35             ~&11.49          \\
	          	~&	Total           ~&32.38       ~&100              ~&81.39		    ~&100	         \\
\midrule[1.0pt]\midrule[1.0pt]
\end{tabular}
\end{table}

\begin{table}[!htbp]
\caption{Strong decays for the $\Upsilon(5S)$ which is assigned to be $Y(10750)$ or $\Upsilon(10860)$ within the screened potential model. }   \label{decay5S}
\begin{tabular}{cccccccccccccccccccccccccc}  \midrule[1.0pt]\midrule[1.0pt]
\multirow{2}{*}{State}
~~&\multirow{2}{*}{Mode}
~~&\multicolumn{2}{c} {\underline{~~~~~~~~~~$Y(10750)$~~~~~~~~~~}}
~~&\multicolumn{2}{c} {\underline{~~~~~~~~~~$\Upsilon(10860)$~~~~~~~~~~}}

\\
~&
~&$\Gamma_{th}$(MeV) ~&$B_{r}(\%)$
~&$\Gamma_{th}$(MeV) ~&$B_{r}(\%)$
\\
\midrule[1.0pt]
$5 ^3S_{1}$	~&	$BB$	        ~&0.20        ~&0.38             ~&7.46	            ~&16.95          \\
           	~&	$BB^*$	        ~&15.63       ~&29.50            ~&20.84            ~&47.34     	 \\
		        ~&	$B^*B^*$	    ~&35.52       ~&67.03            ~&8.26		        ~&18.76	         \\
		        ~&	$B_sB_s$	    ~&1.64        ~&3.09             ~&1.02		        ~&2.31           \\
		        ~&	$B_s^*B_s$	    ~&            ~&                 ~&5.63		        ~&12.79	         \\
		        ~&	$B_s^*B_s^*$	~&            ~&                 ~&0.81             ~&1.84           \\
	          	~&	Total           ~&52.99       ~&100              ~&44.02		    ~&100	         \\
\midrule[1.0pt]\midrule[1.0pt]
\end{tabular}
\end{table}

\subsection{$\Upsilon(10580)$ and $\Upsilon(11020)$}

Taking $\Upsilon(10580)$ and $\Upsilon(11020)$ as the $\Upsilon(4S)$ and $\Upsilon(6S)$ states, respectively, their masses can
be well described in the potential models (see Table~\ref{tabs}). Furthermore, their decay widths can be reasonably understood within the uncertainties (see Table \ref{decay}). However, their dielectron widths are overestimated as the pure $S$-wave states~\cite{Badalian:2009bu,Wang:2018rjg}. To explain the dielectron widths, in Ref.~\cite{Badalian:2009bu}, Badalian \emph{et al.} suggested a $S-D$ mixing in the physical states $\Upsilon(10580)$ and $\Upsilon(11020)$.

We also consider the $\Upsilon(11020)$ resonance as a mixed state via the $\Upsilon(6S)$-$\Upsilon_1(5D)$ mixing. The mixing scheme is adopted as follows:
\begin{equation}
\left(
  \begin{array}{c}
  \Upsilon(M_x) \\
  \Upsilon(11020)\\
  \end{array}\right)=
  \left(
  \begin{array}{cc}
   \cos\theta  &\sin\theta \\
  -\sin\theta  &\cos\theta \\
  \end{array}
\right)
\left(
  \begin{array}{c}
  \Upsilon_1(5D)\\
  \Upsilon(6S)\\
  \end{array}\right).
\end{equation}
The strong decay width of $\Upsilon(11020)$ versus the mixing angle $\theta $ is plotted in Fig.~\ref{6s5d}. It is seen that the total decay width is consistent with experimental data when the mixing angle varies in large range. The current total decay width alone cannot determine the mixing angle. However, the significant leptonic decay width up to $0.130\pm0.030~\rm{keV}$, indicates the $\Upsilon(11020)$ has a large $S-$wave component at least.

From the predicted strong decay properties of $\Upsilon(6S)$ and $\Upsilon_1(5D)$ states listed in Table \ref{decay}, one find that $\Upsilon(6S)$ and $\Upsilon_1(5D)$ states mainly couple to two different channels $BB^*$ and $B^*B^*$, respectively. Furthermore, both $\Upsilon(6S)$ and $\Upsilon_1(5D)$ states are far from the thresholds of $BB^*$ and $B^*B^*$. Thus, the $\Upsilon(6S)$-$\Upsilon_1(5D)$ mixing effect via virtual meson loops may be smaller than that of $\Upsilon(5S)$-$\Upsilon_1(4D)$. If the intermediate meson loops are the main mechanism causing the $S-D$ mixing, the mixing angle for $\Upsilon(6S)$-$\Upsilon_1(5D)$ should be smaller than that for $\Upsilon(5S)$-$\Upsilon_1(4D)$.
To sum up, although the $\Upsilon(6S)$-$\Upsilon_1(5D)$ mixing with a small $5D-$wave component assignment cannot be excluded, we prefer to assign $\Upsilon(11020)$ as the pure $\Upsilon(6S)$ state. In Ref.~\cite{Lu:2016mbb} the authors also expected that there may be less $S-D$ mixing for the $\Upsilon(6S)$ state with the consideration of coupled-channel effects. To better understand the nature of $\Upsilon(11020)$,
the missing $\Upsilon_1(5D)$ is worth looking for in future experiments.
Our predictions of the mass and strong decay widths of $\Upsilon_1(5D)$ may provide helpful information for future experimental observations.

Besides the possibility of $\Upsilon(5S)$-$\Upsilon_1(4D)$ and $\Upsilon(6S)$-$\Upsilon_1(5D)$ mixing, taking the same mixing scheme the decay width of $\Upsilon(10580)$ as $\Upsilon(4S)$-$\Upsilon_1(3D)$ mixing is also shown in Fig.~\ref{4s3d}. The total decay width varies dramatically with the mixing angle, and the zero mixing angle is more favored (actually, we use this case to determine the quark pair creation strength $\gamma$). From Table~\ref{decay}, it is found that the $\Upsilon(4S)$ state mainly couples to the $BB$ channel, and $\Upsilon_1(3D)$ state has strong coupling with the $BB^*$ mode, which suggests that the $\Upsilon(4S)$-$\Upsilon_1(3D)$ mixing effects via virtual meson loops may be negligible. Thus, if the intermediate meson loops are the main mechanism causing the $S-D$ mixing, and the mixing effects can be neglected. It should be mentioned that in Ref.~\cite{Lu:2016mbb} the authors expected that there may sizeable $S-D$ mixing for the $\Upsilon(4S)$ state with the consideration of coupled-channel effects. It should mentioned that the $\Upsilon_1(3D)$ state might be a narrow state with a width of about $20-30$MeV according to our calculations, its decays are governed by the $BB^*$ mode, this decay mode might be suitable to be observed in experiments. Looking for the missing $\Upsilon_1(3D)$ state is useful for better understanding the nature of $\Upsilon(10580)$.


\section{summary}\label{sum}

In this paper, we calculate the spectrum of the higher vector bottomonium sates above the $B\bar{B}$ threshold within both screened and linear
potential models. Then, using the predicted masses and wave functions of these higher vector bottomonium states, their two-body OZI-allowed strong decays are investigated in the $^3P_0$ model.

Combining the productions, mass, and decay width of the higher vector bottomonium states with each other,
we conclude that $Y(10750)$ and $\Upsilon(10860)$ might not be pure $S-$wave and $D-$wave vector bottomonium states.
Then, we further discuss the possibility of $\Upsilon(10860)$ and $Y(10750)$
as mixed states via the $S-D$ mixing. Our results suggest that $Y(10750)$
and $\Upsilon(10860)$ might be mixed states via the $5^3S_1-4^3D_1$ mixing with a sizeable mixing angle $\theta\simeq 20^\circ-30^\circ$.
The components of $Y(10750)$ and $\Upsilon(10860)$ are dominated by the $4^3D_1$ and $5^3S_1$ states, respectively.

Moreover, the strong decay behaviors of the $\Upsilon(10580)$ and $\Upsilon(11020)$ resonances are also discussed.
If the $\Upsilon(10580)$ and $\Upsilon(11020)$ resonances are assigned as the $\Upsilon(4S)$ and $\Upsilon(6S)$ states, respectively,
their observed widths together with masses are consistent with the theoretical predictions.

Finally, it should be mentioned that the mechanism for the $S-D$ mixing is not clear.
If $Y(10750)$ and $\Upsilon(10860)$ as mixed states, several questions should be clarified in future works:
(i) what causes the mixing between the $5^3S_1$ and $4^3D_1$ states; (ii)
and how the masses of the pure $S$- and $D$-wave states are shifted to the physical states by the
configuration mixing.

\section*{  Acknowledgements }

This work is supported by the National Natural Science Foundation of China under Grants No.~11775078, No.~U1832173, No.~11705056, and No.~11405053.



\begin{thebibliography}{99}


\bibitem{Abdesselam:2019gth}
  A.~Abdesselam {\it et al.} [Belle Collaboration],
  Observation of a new structure near 10.75 GeV in the energy dependence of the $e^+e^-\to\Upsilon(nS)\pi^+\pi^-$ ($n=1,2,3$) cross sections,
  arXiv:1905.05521.

\bibitem{Wang:2019veq}
  Z.~G.~Wang,
  Vector hidden-bottom tetraquark candidate: $Y(10750)$,
  arXiv:1905.06610.

\bibitem{Chao:2009}
  B.~Q.~Li and K.~T.~Chao,
  Bottomonium Spectrum with Screened Potential,
  Commun.\ Theor.\ Phys.\  {\bf 52}, 653 (2009).


\bibitem{Meng:2007tk}
  C.~Meng and K.~T.~Chao,
  Scalar resonance contributions to the dipion transition rates of Upsilon(4S,5S) in the re-scattering model,
  Phys.\ Rev.\ D {\bf 77}, 074003 (2008).

\bibitem{Eichten:2007qx}
  E.~Eichten, S.~Godfrey, H.~Mahlke and J.~L.~Rosner,
  Quarkonia and their transitions,
  Rev.\ Mod.\ Phys.\  {\bf 80}, 1161 (2008).

\bibitem{Ebert:2011jc}
  D.~Ebert, R.~N.~Faustov and V.~O.~Galkin,
  Spectroscopy and Regge trajectories of heavy quarkonia and $B_c$ mesons,
  Eur.\ Phys.\ J.\ C {\bf 71}, 1825 (2011).

\bibitem{Ferretti:2013vua}
  J.~Ferretti and E.~Santopinto,
  Higher mass bottomonia,
  Phys.\ Rev.\ D {\bf 90}, 094022 (2014).

\bibitem{Godfrey:2015dia}
  S.~Godfrey and K.~Moats,
  Bottomonium Mesons and Strategies for their Observation,
  Phys.\ Rev.\ D {\bf 92}, 054034 (2015).

\bibitem{Segovia:2016xqb}
  J.~Segovia, P.~G.~Ortega, D.~R.~Entem and F.~Fern¨¢ndez,
  Bottomonium spectrum revisited,
  Phys.\ Rev.\ D {\bf 93}, 074027 (2016).

\bibitem{Deng:2016ktl}
  W.~J.~Deng, H.~Liu, L.~C.~Gui and X.~H.~Zhong,
  Spectrum and electromagnetic transitions of bottomonium,
  Phys.\ Rev.\ D {\bf 95}, 074002 (2017).

\bibitem{Wang:2018rjg}
  J.~Z.~Wang, Z.~F.~Sun, X.~Liu and T.~Matsuki,
  Higher bottomonium zoo,
  Eur.\ Phys.\ J.\ C {\bf 78}, 915 (2018).


\bibitem{Wei-Zhao:2013sta}
  T.~Wei-Zhao, C.~Lu, Y.~You-Chang and C.~Hong,
  Bottomonium states versus recent experimental observations in the QCD-inspired potential model,
  Chin.\ Phys.\ C {\bf 37}, 083101 (2013).

\bibitem{Badalian:2009bu}
  A.~M.~Badalian, B.~L.~G.~Bakker and I.~V.~Danilkin,
  Dielectron widths of the S-, D-vector bottomonium states,
  Phys.\ Atom.\ Nucl.\  {\bf 73}, 138 (2010)

\bibitem{Badalian:2008ik}
  A.~M.~Badalian, B.~L.~G.~Bakker and I.~V.~Danilkin,
  On the possibility to observe higher $n^3 D_1$ bottomonium states in the $e^+ e^-$ processes,
  Phys.\ Rev.\ D {\bf 79}, 037505 (2009).


\bibitem{Ali:2009es}
  A.~Ali, C.~Hambrock and M.~J.~Aslam,
  A Tetraquark interpretation of the BELLE data on the anomalous $\Upsilon(1S) \pi^+\pi^-$ and $\Upsilon(2S) \pi^+\pi^-$ production near the $\Upsilon(5S)$ resonance, Phys.\ Rev.\ Lett.\  {\bf 104}, 162001 (2010);
  Erratum: [Phys.\ Rev.\ Lett.\  {\bf 107}, 049903 (2011)].


\bibitem{Dubynskiy:2008mq}
  S.~Dubynskiy and M.~B.~Voloshin,
  Hadro-Charmonium,
  Phys.\ Lett.\ B {\bf 666}, 344 (2008).

\bibitem{Eichten:2004uh}
  E.~J.~Eichten, K.~Lane and C.~Quigg,
  Charmonium levels near threshold and the narrow state $X(3872) \to \pi^{+}\pi^{-}J/\psi$,
  Phys.\ Rev.\ D {\bf 69}, 094019 (2004).

\bibitem{Eichten:2005ga}
  E.~J.~Eichten, K.~Lane and C.~Quigg,
  New states above charm threshold,
  Phys.\ Rev.\ D {\bf 73}, 014014 (2006)
  Erratum: [Phys.\ Rev.\ D {\bf 73}, 079903 (2006)].

\bibitem{Rosner:2004wy}
  J.~L.~Rosner,
  $\Psi''$ decays to charmless final states,
  Annals Phys.\  {\bf 319}, 1 (2005).

\bibitem{Close:2006gr}
  F.~E.~Close, C.~E.~Thomas, O.~Lakhina and E.~S.~Swanson,
  Canonical interpretation of the $D_{sJ}(2860)$ and $D_{sJ}(2690)$,
  Phys.\ Lett.\ B {\bf 647}, 159 (2007).

\bibitem{Zhong:2009sk}
  X.~H.~Zhong and Q.~Zhao,
  Strong decays of newly observed $D_{sJ}$ states in a constituent quark model with effective Lagrangians,
  Phys.\ Rev.\ D {\bf 81}, 014031 (2010).

\bibitem{Zhong:2010vq}
  X.~H.~Zhong,
  Strong decays of the newly observed $D(2550)$, $D(2600)$, $D(2750)$, and $D(2760)$,
  Phys.\ Rev.\ D {\bf 82}, 114014 (2010).

\bibitem{Li:2010vx}
  D.~M.~Li, P.~F.~Ji, and B.~Ma,
  The newly observed open-charm states in quark model,
  Eur.\ Phys.\ J.\ C {\bf 71}, 1582 (2011).


\bibitem{Li:2009qu}
  D.~M.~Li and B.~Ma,
  Implication of BaBar's new data on the $D_{s1}(2710)$ and $D_{sJ}(2860)$,
  Phys.\ Rev.\ D {\bf 81}, 014021 (2010).

\bibitem{Chen:2011rr}
  B.~Chen, L.~Yuan and A.~Zhang,
  Possible 2S and 1D charmed and charmed-strange mesons,
  Phys.\ Rev.\ D {\bf 83}, 114025 (2011).

\bibitem{Chen:2016spr}
  H.~X.~Chen, W.~Chen, X.~Liu, Y.~R.~Liu and S.~L.~Zhu,
  A review of the open charm and open bottom systems,
  Rept.\ Prog.\ Phys.\  {\bf 80}, 076201 (2017).


\bibitem{Liu:2019zuc}
  M.~S.~Liu, Q.~F.~L\"u, X.~H.~Zhong and Q.~Zhao,
  Fully-heavy tetraquarks,
  arXiv:1901.02564.


\bibitem{Eichten:1974af}
  E.~Eichten, K.~Gottfried, T.~Kinoshita, J.~B.~Kogut, K.~D.~Lane and T.~M.~Yan,
  The Spectrum of Charmonium,
  Phys.\ Rev.\ Lett.\  {\bf 34}, 369 (1975)
  Erratum: [Phys.\ Rev.\ Lett.\  {\bf 36}, 1276 (1976)].


\bibitem{Eichten:1978tg}
  E.~Eichten, K.~Gottfried, T.~Kinoshita, K.~D.~Lane and T.~M.~Yan,
  Charmonium: The Model,
  Phys.\ Rev.\ D {\bf 17}, 3090 (1978)
  Erratum: [Phys.\ Rev.\ D {\bf 21}, 313 (1980)].

\bibitem{Godfrey:1985xj}
  S.~Godfrey and N.~Isgur,
  Mesons in a Relativized Quark Model with Chromodynamics,
  Phys.\ Rev.\ D {\bf 32}, 189 (1985).


\bibitem{Barnes:2005pb}
  T.~Barnes, S.~Godfrey and E.~S.~Swanson,
  Higher charmonia,
  Phys.\ Rev.\ D {\bf 72}, 054026 (2005).

\bibitem{Li:2009zu}
  B.~Q.~Li and K.~T.~Chao,
  Higher Charmonia and X,Y,Z states with Screened Potential,
  Phys.\ Rev.\ D {\bf 79}, 094004 (2009).



\bibitem{ChaoKT93}
Y. B. Ding, K. T. Chao and D. H. Qin,
Screened $Q\bar{Q}$ potential and spectrum of heavy quarkonium,
Chin. Phys. Lett.{\bf 10}, 460 (1993).

\bibitem{Liu:2011yp}
  J.~F.~Liu and G.~J.~Ding,
  Bottomonium Spectrum with Coupled-Channel Effects,
  Eur.\ Phys.\ J.\ C {\bf 72}, 1981 (2012).

\bibitem{Lu:2016mbb}
  Y.~Lu, M.~N.~Anwar and B.~S.~Zou,
  Coupled-Channel Effects for the Bottomonium with Realistic Wave Functions,
  Phys.\ Rev.\ D {\bf 94}, 034021 (2016).

\bibitem{Haicai}
 Chong-Hai Cai and Lei Li,
  Radial equation of bound state and binding energies of $\Xi^-$ hypernuclei,
  Chin. Phys. C {\bf 27}, 1005 (2003).

\bibitem{Deng:2016stx}
  W.~J.~Deng, H.~Liu, L.~C.~Gui and X.~H.~Zhong,
  Charmonium spectrum and their electromagnetic transitions with higher multipole contributions,
  Phys.\ Rev.\ D {\bf 95}, 034026 (2017).

\bibitem{Micu:1968mk}
  L.~Micu,
  Decay rates of meson resonances in a quark model,
  Nucl.\ Phys.\ B{\bf10}, 521 (1969).

\bibitem{LeYaouanc:1972vsx}
  A.~Le Yaouanc, L.~Oliver, O.~Pene, and J.~C.~Raynal,
  Naive quark pair creation model of strong interaction vertices,
  Phys.\ Rev.\ D {\bf 8}, 2223 (1973).

\bibitem{LeYaouanc:1973ldf}
  A.~Le Yaouanc, L.~Oliver, O.~Pene, and J.-C.~Raynal,
  Naive quark pair creation model and baryon decays,
  Phys.\ Rev.\ D {\bf 9}, 1415 (1974).



\bibitem{Jacob:1959at}
  M.~Jacob and G.~C.~Wick,
  On the general theory of collisions for particles with spin,
  Ann. Phys. (N.Y.)\  {\bf 7}, 404 (1959)
  ;\  {\bf 281}, 774 (2000).

\bibitem{Gui:2018rvv}
  L.~C.~Gui, L.~S.~Lu, Q.~F.~L\"u, X.~H.~Zhong and Q.~Zhao,
  Strong decays of higher charmonium states into open-charm meson pairs,
  Phys.\ Rev.\ D {\bf 98}, 016010 (2018).

\bibitem{Li:2019tbn}
  Q.~Li, M.~S.~Liu, L.~S.~Lu, Q.~F.~L\"u, L.~C.~Gui and X.~H.~Zhong,
  Excited bottom-charmed mesons in a nonrelativistic quark model,
  Phys.\ Rev.\ D {\bf 99}, 096020 (2019).


\bibitem{Lu:2016bbk}
  Q.~F.~L\"{u}, T.~T.~Pan, Y.~Y.~Wang, E.~Wang, and D.~M.~Li,
  Excited bottom and bottom-strange mesons in the quark model,
  Phys.\ Rev.\ D {\bf 94}, 074012 (2016).


\bibitem{Tanabashi:2018oca}
  M.~Tanabashi {\it et al.} [Particle Data Group],
  Review of Particle Physics,
  Phys.\ Rev.\ D {\bf 98}, 030001 (2018).

\bibitem{Abdesselam:2016tbc}
  A.~Abdesselam {\it et al.},
  Study of Two-Body $e^+e^- \to B_s^{(*)}\bar{B}_s^{(*)}$ Production in the Energy Range from 10.77 to 11.02 GeV,
  arXiv:1609.08749 [hep-ex].


\bibitem{Wu:2011yb}
  X.~G.~Wu and Q.~Zhao,
  The mixing of $D_{s1}(2460)$ and $D_{s1}(2536)$,
  Phys.\ Rev.\ D {\bf 85}, 034040 (2012).

\bibitem{Cao:2017lui}
  Z.~Cao and Q.~Zhao,
  Impact of $S$-wave thresholds $D_{s1}\bar{D}_{s}+c.c.$ and $D_{s0}\bar{D}^*_{s}+c.c.$ on vector charmonium spectrum,
  Phys.\ Rev.\ D {\bf 99}, 014016 (2019).

\end{thebibliography}
\end{document}